\documentclass[preprint,preprintnumbers,nofootinbib]{revtex4}

\usepackage[dvips]{color}% Use color

\newcommand{\diag}{{\rm diag}}

\begin{document}
\vspace{2cm}
\preprint{OU-HET 616/2008}

\title{Tri-Bimaximal Mixing from Twisted Friedberg-Lee Symmetry}
\author{Takeshi Araki$^a$}
\author{Ryo Takahashi$^b$}

\affiliation{
$^a$Department of Physics, National Tsing Hua University,
Hsinchu, Taiwan 300\\
$^b$Department of Physics, Graduate School of Science, Osaka University, 
Toyonaka, Osaka 560-0043, Japan}

\begin{abstract}
We investigate the Friedberg-Lee (FL) symmetry and 
its promotion to include the $\mu - \tau$ symmetry, and call that 
the twisted FL symmetry. 
Based on the twisted FL symmetry, two possible schemes are presented 
toward the realistic neutrino mass spectrum and the tri-bimaximal mixing. 
In the first scheme, we suggest the semi-uniform translation of the FL symmetry.
The second one is based on the $S_3$ permutation family symmetry. 
The breaking terms, which are twisted FL symmetric, are introduced. 
Some viable models in each scheme are also presented.
\end{abstract}

\maketitle

\section{Introduction}
The precision measurements of the neutrino oscillation have suggested that 
there are large mixings among three generations in the lepton sector unlike 
the quark sector. The current experimental data of mixing angles \cite{exp} is 
well approximated by the tri-bimaximal mixing (TBM) \cite{TB}, which is given by
\begin{eqnarray}
 V_{TB}=
  \left(\begin{array}{ccc}
	  2/\sqrt{6} & 1/\sqrt{3} & 0 \\
	  -1/\sqrt{6} & 1/\sqrt{3} & -1/\sqrt{2} \\
	  -1/\sqrt{6} & 1/\sqrt{3} & 1/\sqrt{2}
  \end{array}\right)\ .\label{eq:TB}
\end{eqnarray}
The properties of this mixing matrix are that the second generation of the 
neutrino mass eigenstate is represented by trimaximal mixture of all flavor 
eigenstates, $\nu_2=\sum_\alpha\nu_\alpha/\sqrt{3}$, and the third generation 
is bimaximal mixture of $\mu$ and $\tau$ neutrinos, 
$\nu_3=(-\nu_\mu+\nu_\tau)/\sqrt{2}$, in the diagonal basis of the charged 
leptons.
Naively, it appears that Eq. (\ref{eq:TB}) implies the following forms of the 
neutrino mass matrix in the flavor basis,
\begin{eqnarray}
 {\cal M}^{\nu}=
 \frac{m_1}{6}
 \left(\begin{array}{ccc}
  4 & -2 & -2 \\
  -2 & 1 & 1 \\
  -2 & 1 & 1
 \end{array}\right)
 +\frac{m_2}{3}
 \left(\begin{array}{ccc}
  1 & 1 & 1 \\
  1 & 1 & 1 \\
  1 & 1 & 1
 \end{array}\right)
 +\frac{m_3}{2}
 \left(\begin{array}{ccc}
  0 & 0 & 0 \\
  0 & 1 & -1 \\
  0 & -1 & 1
 \end{array}\right)\ ,\label{eq:Mnu}
\end{eqnarray}
where $m_i$ $(i=1\sim3)$ are the neutrino mass eigenvalues. 

Such suggestive forms of the generation mixing and the neutrino mass matrices 
give us a strong motivation to look for a flavor structure of the lepton 
sector. In fact there exist $\mu - \tau$ permutation symmetry in Eq. 
(\ref{eq:Mnu}). The Majorana mass matrix having the $\mu - \tau$ 
symmetry is written by \cite{mutau}
\begin{eqnarray}
{\cal M}^{\nu}=
\left(
\begin{array}{ccc}
A & B & B \\
B & C & D \\
B & D & C \\
\end{array}
\right),
\label{eq:mutau}
\end{eqnarray}
where we assume that $A,B,C$ and $D$ are real.
This form of the neutrino mass matrix in the diagonal basis of the charged 
leptons can be diagonalized by the following mixing matrix,
\begin{eqnarray}
V=
\left(
\begin{array}{ccc}
\cos\theta_{12}           & \sin\theta_{12}          & 0          \\
-\sin\theta_{12}/\sqrt{2} & \cos\theta_{12}/\sqrt{2} & 1/\sqrt{2} \\
-\sin\theta_{12}/\sqrt{2} & \cos\theta_{12}/\sqrt{2} & -1/\sqrt{2}
\end{array}
\right).
\label{eq:bm}
\end{eqnarray}
This mixing matrix suggests that $\theta_{23}$ is maximal $\pi/4$, 
$\theta_{13}$ is vanishing and $\theta_{12}$ is undetermined. When there is a 
certain relation among the mass parameters in the matrix Eq. (\ref{eq:mutau}), 
which is $A+B=C+D$, the $\sin\theta_{12}$ is determined such as 
$\sin\theta_{12}=1/\sqrt{3}$ and thus the mixing matrix becomes the 
TBM one. 
Indeed, each mass matrix in Eq. (\ref{eq:Mnu}) possesses the relation $A+B=C+D$
as well as the $\mu-\tau$ symmetry.
What is the origin of such a special texture? A number of proposals 
based on a flavor symmetry to unravel it have been elaborated \cite{models}.

Recently, Friedberg and Lee have proposed a new type of family symmetry 
called Friedberg-Lee (FL) symmetry \cite{FL}.
The FL symmetry is a translational hidden family symmetry and predicts 
one massless neutrino and the
trimaximal mixture for the second generation neutrino.
Several possible origins of the FL symmetry have been discussed in Ref. \cite{gFL}. 
In addition to the symmetry, if we further impose the $\mu-\tau$ symmetry, 
the model leads to the TBM \cite{Xing,Xing2} and 
can partly realize the structure of Eq. (\ref{eq:Mnu}).
That is, the FL symmetry can be the origin of the condition $A+B=C+D$.
However, the model cannot reproduce the realistic neutrino mass spectrum 
as we shall explain later.
Therefore, in order to construct a viable flavor model based on the FL 
symmetry, some breakings of the symmetry, additional symmetries and/or 
extensions are needed. 
%In fact a simple scheme of breaking such as 
%$\nu_e\rightarrow\kappa^\ast\nu_e$ with $\kappa\neq1$ has been proposed.

In this letter, we consider the $\mu-\tau$ symmetric extension of the FL symmetry 
and refer to it as the twisted FL symmetry.
Then we propose two possible schemes to obtain the experimentally 
favored neutrino mass spectrum and the TBM.
The first scheme is based on the non-uniformity of the FL translation 
and its breaking.
In the second scheme, we introduce the $S_3$ permutation symmetry as a family one.
It is well known that the second term in Eq. (\ref{eq:Mnu}), which is called 
a democratic form, can be realized by the $S_3$ symmetry \cite{s3,s31}. 
The other terms of Eq. (\ref{eq:Mnu}) can be generated by introducing the $S_3$ 
symmetry breaking terms which are governed by the twisted FL symmetry.
In both schemes, the $\mu-\tau$ symmetry remains intact and the FL symmetry 
(and $S_3$ symmetry) ensures the condition $A+B=C+D$ mentioned above.

The letter is organized as follows. 
In Section II, we define the twisted FL symmetry and propose a realistic model 
for the Majorana neutrinos on the basis of the non-unifrom translation.
In Section III, we focus on the $S_3$ permutation symmetry and propose models 
for both Dirac and Majorana neutrinos.
Section IV is devoted to a summary. 
Detailed derivations of some neutrino mass 
matrices based on the twisted FL symmetry are summarized in Appendix A.

\section{Twisted FL Symmetry and Uniformity Breaking}
We start with the ordinary (type-I) seesaw mechanism \cite{seesaw} with the 
three right-handed heavy Majorana neutrinos. After integrating out the 
right-handed Majorana neutrinos, the mass terms of the charged leptons and 
effective light Majorana neutrinos are written as
\begin{eqnarray}
 -{\cal L}=
  \bar{\ell}_{Li} M_{ij}^e \ell_{Rj}
  +\overline{\nu^c}_{i}{\cal M}_{ij}^{\nu}\nu_{j}+h.c.\ ,
\end{eqnarray}
where $\ell_{Li}$ and $\ell_{Ri}$ are the left- and right-handed charged 
leptons, and $\nu_i$ are the light Majorana neutrinos. 
The $i$ and $j$ stand for family indices. 
The charged lepton and effective Majorana neutrino mass matrices are 
denoted as $M^e$ and ${\cal M}^\nu$, respectively.
In what follows, we take ${\cal M}^\nu$ as a real 
matrix since we do not discuss CP violation in this letter and the 
following discussions can be simply implied to a complex case.
%without changing results.
In the diagonal basis of the charged leptons, ${\cal M}^{\nu}$ is diagonalized by 
the Maki-Nakagawa-Sakata (MNS) matrix: 
$V_{MNS}^{\dag}\ {\cal M}^{\nu}\ V_{MNS}^{*}=\diag (m_1,m_2,m_3)$. 
In this basis, we impose the following translational family symmetry as a hidden one 
only on the light Majorana neutrino mass term,
\begin{eqnarray}
 \nu_i \rightarrow \nu_i^{'}=S_{ij}\nu_j + \eta_i \xi\ ,\label{eq:TFL}
\label{uni}
\end{eqnarray}
where $\xi$ is a space-time independent Grassmann parameter, $\xi^2 =0$, 
$\eta = (\eta_1,\eta_2,\eta_3)^T$ are c-numbers, and 
$S$ is the permutation matrix between the 
second and third families:
\begin{eqnarray}
 S=
 \left(\begin{array}{ccc}
 1 & 0 & 0 \\
 0 & 0 & 1 \\
 0 & 1 & 0
 \end{array}\right)\ . 
\end{eqnarray}
This kind of family symmetry has been proposed by Friedberg and Lee in 
Ref. \cite{FL} for the first time, and is called Friedberg-Lee (FL) symmetry. 
In Eq. (\ref{eq:TFL}), we combine the FL symmetry with the $\mu-\tau$ 
one and call it the twisted FL symmetry from now on.

In the following subsections, we propose two neutrino-flavor models on the basis of 
the twisted FL symmetry. 
The first model corresponds to a naive $\mu - \tau$ symmetric 
extension of the original FL symmetry. We show that the model can partially 
realize the structure of Eq. (\ref{eq:Mnu}) but should be improved to obtain 
the experimentally favored neutrino mass spectrum. 
Then we 
%suggest the possible modification and 
propose a realistic model.
 
\subsection{Uniform translation}
First, we consider the uniform translation, that is $\eta_1 = 
\eta_2 = \eta_3$ in Eq. (\ref{eq:TFL}). Such a translation leads to the 
following effective Majorana neutrino mass matrix,\footnote{A detailed 
derivation is given in Appendix A.}
\begin{eqnarray}
  {\cal M}^{\nu}=
 \frac{B}{2}
 \left(\begin{array}{ccc}
  4 & -2 & -2 \\
  -2 & 1 & 1 \\
  -2 & 1 & 1
 \end{array}\right)
 +\left(A+\frac{B}{2}\right)
 \left(\begin{array}{ccc}
  0 & 0 & 0 \\
  0 & 1 & -1 \\
  0 & -1 & 1
 \end{array}\right)\ .\label{eq:eFL}
\end{eqnarray}
We find that the first and third terms in Eq. (\ref{eq:Mnu}) are obtained.
The mass matrix can be diagonalized by the TBM matrix Eq. (\ref{eq:TB})
and leads to the neutrino masses
\begin{eqnarray}
 m_1 = 3B\ ,\ m_2 = 0\ ,\ m_3 = 2A+B\ .
\end{eqnarray}
Since $m_2 > m_1$ is suggested from the solar neutrino oscillation
 experiment, this model is inconsistent with the experimental data.
In order to obtain the proper neutrino mass spectrum, we need to introduce the symmetry breaking terms in Eq. (\ref{eq:eFL}).
However, it seems hard to derive $m_2 > m_1$ in this model because
it may be natural to assume that the magnitude of the breaking 
terms should be smaller than that of the terms based on a symmetry.
In that sense, this model cannot reproduce a realistic neutrino mass spectrum.

\subsection{Semi-uniform translation}
It has been shown that the twisted FL symmetry with the uniform 
translation cannot give the proper neutrino mass spectrum.
Here, we consider a semi-uniform translation, $\eta_2=\eta_3$ and 
$\sum\eta_i=0$ $(i=1\sim3)$.\footnote{This translation, equivalently 
$\eta_i\propto(2,-1,-1)$, corresponds to the case of $\kappa=-1/2$ in Ref. 
\cite{Xing2}.}
This translation leads to the effective Majorana mass matrix 
as\footnote{A detailed derivation is also given in Appendix A.}
\begin{eqnarray}
 {\cal M}^{\nu}=
 \frac{B}{2}
 \left(\begin{array}{ccc}
  1 & 1 & 1 \\
  1 & 1 & 1 \\
  1 & 1 & 1
 \end{array}\right)
 +\left( A+\frac{B}{2} \right)
 \left(\begin{array}{ccc}
  0 & 0 & 0 \\
  0 & 1 & -1 \\
  0 & -1 & 1
 \end{array}\right)\ .\label{eq:eFL2}
\end{eqnarray}
We find that the second and third terms of Eq. (\ref{eq:Mnu}) are obtained. 
This mass matrix can be diagonalized by the TBM matrix and
leads to a massless neutrino, 
$m_1=0$.
%\footnote{It is remarked that the 
%translation $\eta_i \propto(0,-1,1)$ which corresponds to the third column of 
%Eq. (\ref{eq:TB}) results in the nearly TBM and $m_3=0$.} 
 That is, we can easily realize a viable model in terms of the 
twisted FL symmetry with the semi-uniform translation.

The nonzero $m_1$ can be also obtained by introducing the symmetry
 breaking terms in Eq. (\ref{eq:eFL2}). It is seen that if the 
breaking term has a particular structure such as the first term in Eq. (\ref{eq:Mnu}),
\begin{eqnarray}
 c\left(\begin{array}{ccc}
  4 & -2 & -2 \\
  -2 & 1 & 1 \\
  -2 & 1 & 1
 \end{array}\right), \label{eq:eFL4}
\end{eqnarray}
the TBM and a massive $m_1$ are realized. Actually, such a form of 
a mass matrix can be obtained in the twisted FL scheme with non-uniform 
translation, $\eta_i\neq \eta_j$ $(i\neq j)$ with 
$3\eta_1=\sum\eta_i$.\footnote{A detailed derivation is given in Appendix A.}
 Consequently, the $\mu-\tau$ symmetry and the condition $A+B=C+D$ mentioned in
 the introduction are remained. They are just the result of the twisted FL 
operation.

As a result, the whole mass matrix takes the same from as Eq. (\ref{eq:Mnu}), 
%which can be diagonalized by the TBM matrix, 
and neutrino masses are written as
\begin{eqnarray}
 m_1 = 6c\ ,\ m_2 = \frac{3}{2}B\ ,\ 
 m_3 = 2A+B\ .
\end{eqnarray}
The neutrino mass spectrum is the normal hierarchy.
The smallness of $m_1$ is easily understood by the fact that the mass is 
induced from the symmetry breaking.

\section{Democratic texture with twisted FL symmetry}
In this section, we propose other realizations of the 
experimentally favored neutrino mass spectrum and the TBM in terms of the 
twisted FL symmetry. We take the $S_3$ symmetry as a 
fundamental flavor symmetry and introduce breaking terms which are twisted FL 
symmetric.

\subsection{Dirac neutrinos}
We focus on the Dirac neutrinos in this subsection. 
The mass terms of the charged leptons and 
Dirac neutrinos are written as
\begin{eqnarray}
 -{\cal L}_{\ell_M}=
  \bar{\ell}_{Li} M_{ij}^e \ell_{Rj}
  +\bar{\nu}_{Li} M_{ij}^D \nu_{Rj}+h.c.\ ,
\end{eqnarray}
where $\nu_{Li}$ and $\nu_{Ri}$ are the left- and right-handed Dirac neutrinos, 
respectively.
The Dirac mass matrix is denoted as $M^D$, and for simplicity we assume that 
$M^D$ is real. 
Here we separately impose the $S_3$ permutation symmetry for the 
left- and right-handed neutrinos as a flavor symmetry. They are 
shown by $S_{3L}$ and $S_{3R}$, respectively. It is well known that the 
minimal introduction of matter contents in the Higgs sector, which contains a 
single elementary scalar in the singlet representation of both $S_{3L}$ and 
$S_{3R}$ symmetries, leads to the following invariant Dirac mass matrix under 
$S_{3L}\times S_{3R}$ symmetry,\footnote{By changing the basis of Eq. 
(\ref{demo}), the mass matrix has only the 3-3 element. Both the mass matrix 
and one in Eq. (\ref{demo}) have only one non-vanishing mass eigenvalue and the
 same mass spectrum. However, the flavor mixings are different. The mass matrix
 containing only the 3-3 element does not lead to any mixings clearly but the 
democratic form of matrix in Eq. (\ref{demo}) has large mixings. This has been 
pointed out in Ref. \cite{Haba}. In this letter, we take the democratic basis 
in Eq. (\ref{demo}) for phenomenological interests.} 
\begin{eqnarray}
 M^D=C
  \left(\begin{array}{ccc}
	  1 & 1 & 1 \\
	  1 & 1 & 1 \\
	  1 & 1 & 1
  \end{array}\right)\ .
\label{demo}
\end{eqnarray}
By comparing the mass matrix with Eq. (\ref{eq:Mnu}), one can see that the 
resulting mass spectrum is $m_2 = 3C$ and $m_1 = m_3 = 0$ with the TBM. 
Thus the model is experimentally unfavored. 
The possible way to rescue the model is to obtain the non-vanishing $m_1$ or 
$m_3$. We consider a case to have non-zero $m_1$ by introducing the symmetry 
breaking term based on the FL symmetry in what 
follows.\footnote{Some proposals of breaking of such a $S_3$ 
flavor symmetry, which leads to the democratic mass matrix, have been discussed
 in model including extended Higgs sector \cite{s3}, $O(3)$ breaking mechanism 
\cite{Tanimoto:1999pj}, and considering perturbative effects \cite{Wolf}.}

Before introducing the breaking terms, we consider an extension of the 
twisted FL symmetry for the Dirac neutrinos. We extend the 
twisted FL symmetry for the Dirac neutrinos in the following way,
\begin{eqnarray}
&&\nu_{Li} \rightarrow \nu_{Li}^{'}=S_{ij}^L\nu_{Lj}+\xi_L\ ,\label{eq:LFL}\\
&&\nu_{Ri} \rightarrow \nu_{Ri}^{'}=S_{ij}^R\nu_{Rj}+\xi_R\ .\label{eq:RFL}
\end{eqnarray}
Under these transformations, the form of the Dirac mass matrix is constrained 
to the one given in Eq. (\ref{eq:eFL4}) as discussed in Appendix A. 
Note that the Dirac mass matrix is 
invariant under the independent transformations for the left- and right-handed 
neutrinos such as $(\nu_{Li},\nu_{Ri})\rightarrow(\nu_{Li}',\nu_{Ri})$ or 
$(\nu_{Li},\nu_{Ri}')$.\footnote{If we impose the same twisted FL symmetry on 
both left- and right-handed neutrinos, 
$\nu_{Li,Ri}\rightarrow\nu_{Li,Ri}'=S_{ij}\nu_{Lj,Rj}+\xi$, the 
form of the mass matrices given in Eq. (\ref{eq:eFL}) is obtained as the Dirac 
neutrino mass matrix.} For the Dirac neutrinos, the twisted FL symmetry each 
for the left- and right-handed neutrinos can lead to the strange form of mass 
matrix given in Eq. (\ref{eq:eFL4}) as preserving the uniformity. That is one 
of the interesting features of the twisted FL symmetry for the Dirac neutrinos. 

We add the mass matrix Eq. (\ref{eq:eFL4}) to Eq. (\ref{demo}) 
as the breaking terms for the $S_{3L}\times S_{3R}$ flavor symmetry. 
Consequently, the total Dirac neutrino mass matrix is written as
\begin{eqnarray}
 M^D=C
 \left(\begin{array}{ccc}
	  1 & 1 & 1 \\
	  1 & 1 & 1 \\
	  1 & 1 & 1
  \end{array}\right)+c
 \left(\begin{array}{ccc}
     4 & -2 & -2 \\
     -2 & 1 & 1 \\
     -2 & 1 & 1
  \end{array}\right)\ ,\label{eq:MD}
\end{eqnarray}
and the neutrino mass spectrum is given by
\begin{eqnarray}
 m_1 = 6c,\ \ m_2 = 3C,\ \ m_3 = 0\ ,
\end{eqnarray}
which suggests the inverted neutrino mass hierarchy. 
The TBM can be also realized. 
This is a minimal scheme: the twisted FL symmetry for 
the Dirac neutrinos is introduced as the symmetry of the breaking term.
We note that the whole mass matrix Eq. (\ref{eq:MD}) violates the 
$S_{3L}\times S_{3R}$ symmetry but still preserves the $\mu-\tau$ symmetry.

\subsection{Majorana neutrinos}
Here a similar model is presented for the Majorana neutrinos. The most general 
Majorana mass matrix based on the $S_3$ symmetry can be written as  
\begin{eqnarray}
 {\cal M}^{\nu}=
  \left(\begin{array}{ccc}
	  E & F & F \\
	  F & E & F \\
	  F & F & E
  \end{array}\right)\ .
\end{eqnarray}
This matrix has two degenerate eigenvalues,
\begin{eqnarray}
 m_1 = E-F,\ \ m_2 = E+2F,\ \ m_3 = E-F\ ,
\end{eqnarray}
where we take $m_1=m_3$ and assume that ${\cal M}^\nu $ is real. 
To obtain the mass difference between $m_1$ and $m_3$,
 we introduce Eq. (\ref{eq:eFL}) as a symmetry breaking term. Then the total 
Majorana neutrino mass matrix is written as
\begin{eqnarray}
 {\cal M}^\nu =
  \left(\begin{array}{ccc}
	  E & F & F \\
	  F & E & F \\
	  F & F & E
  \end{array}\right)
  +\frac{b}{2}
  \left(\begin{array}{ccc}
     4 & -2 & -2 \\
     -2 & 1 & 1 \\
     -2 & 1 & 1
  \end{array}\right)
 +\left( a + \frac{b}{2} \right)
  \left(\begin{array}{ccc}
     0 & 0 & 0 \\
     0 & 1 & -1 \\
     0 & -1 & 1
  \end{array}\right) .
\end{eqnarray}
This matrix can also be diagonalized by the TBM matrix, and mass eigenvalues
are given by
\begin{eqnarray}
 m_1 = E-F+3b,\ \ m_2 = E+2F,\ \ m_3 = m_1 + 2(a+b)\ .
\end{eqnarray}
It seems that a natural mass spectrum induced 
by this model is the quasi-degenerate one because the mass difference
 between $m_1$ and $m_3$ is determined by only the small symmetry breaking 
parameters $a$ and $b$. 
For instance, in accordance with Ref. \cite{Wolf}, if 
we require $F = -2E$ and $E>0$, the breaking parameters must be $b<0$ and 
$|a| \gg |b|$ to satisfy the recent neutrino oscillation update given in Ref. \cite{exp}
\begin{eqnarray}
 &&\Delta m_{21}^2
  = (7.695 \pm 0.645)\times 10^{-5}\ {\rm eV}^2\ ,\\
 &&\Delta m_{31}^2
  = (2.40^{+0.12}_{-0.11})\times 10^{-3}\ {\rm eV}^2\ .
\end{eqnarray}
Then this model results in the degenerate mass spectrum.

\section{Summary}
We have considered the 
promotion of the FL symmetry to contain the 
$\mu - \tau$ symmetry and call this the twisted FL symmetry.
It has also been shown that the symmetry with the uniform translation for 
neutrinos can partially realize the desired forms of the neutrino mass matrix for 
the TBM but should be improved to obtain the experimentally 
favored neutrino mass spectrum.
Then, we have presented two possible schemes which can achieve a realistic 
neutrino mass spectrum and the TBM.

The first scheme is based on the non-uniformity of the FL symmetry. 
We have discussed that for Majorana neutrinos and presented 
a viable model which predicts the normal mass hierarchy.
In the second scheme, we have focused on the $S_3$ permutation family symmetry. 
The breaking terms as preserving the twisted FL symmetry have been introduced there. 
We have examined that for both Dirac and Majorana neutrinos. 
In the case of the Dirac neutrinos, 
%the twisted FL symmetry has been independently 
%imposed on the left- and right-handed neutrinos. 
the model predicts the massless third generation neutrino and the inverted 
mass hierarchy. 
The model for the Majorana neutrinos in this scheme results in the degenerate 
mass spectrum.
All of the models lead to the TBM.

\acknowledgements
One of authors (T.A.) would like to thank Prof. C. Q. Geng for useful 
discussions. The work of R.T. has been supported by the Japan Society of 
Promotion of Science. The authors wish to thank the Taiwan National Center for 
Theoretical Sciences, where this work was initiated during the NCTS workshop 
``Summer Institute 2008''. 

\appendix
\section{Twisted Friedberg-Lee symmetry}
In this appendix, we give a detailed discussion how to obtain the neutrino 
mass matrices used in the main part from the twisted FL symmetry. 
Discussions are presented for both Majorana and Dirac neutrinos.

\subsection{Majorana neutrinos}
Let us consider the Majorana neutrino case and define the mass term as follows,
\begin{eqnarray}
 -{\cal L}_{\nu}=\bar{\nu}^c_{i} {\cal M}_{ij}^{\nu} \nu_{j}\ .
\end{eqnarray}
After operating the twisted FL transformation given in Eq. (\ref{eq:TFL}) on 
the neutrino fields, the mass term becomes
\begin{eqnarray}
 \bar{\nu}^c_{i} {\cal M}_{ij}^{\nu} \nu_{j}
 &\rightarrow&
 \left[\ \bar{\nu}^c_{k}S_{ki}+\bar{\eta}_i\xi\ \right]
 {\cal M}^{\nu}_{ij}\left[\ S_{jl}\nu_{l}+\eta_j \xi \ \right]\ .
\end{eqnarray}
Hence, the invariance of Eq. (\ref{eq:TFL}) requires the $\mu  - \tau$ symmetry, 
$S{\cal M}^{\nu}S={\cal M}^\nu$, and the translational symmetry,
$\bar{\eta}{\cal M^\nu}={\cal M^\nu\eta}=0$.
Firstly, the $\mu  - \tau$ symmetry restricts the form of ${\cal M}^\nu$ to that of 
Eq. (\ref{eq:mutau}).
For convenience we adopt the following parameterization here 
\begin{eqnarray}
 {\cal M}^{\nu}=
 \left(\begin{array}{ccc}
 D & -2C & -2C \\
 -2C & A+B & -A \\
 -2C & -A & A+B
 \end{array}\right)\ .
\end{eqnarray}
Secondly, from the translational symmetry, one finds the conditions 
\begin{eqnarray}
 {\cal M}^{\nu}_{ij}\ \eta_{j}=
 \left(\begin{array}{ccc}
 D & -2C & -2C \\
 -2C & A+B & -A \\
 -2C & -A & A+B
 \end{array}\right)
 \left(\begin{array}{c}
 \eta_{1} \\ \eta_{2} \\ \eta_{3}
 \end{array}\right)=0\ .
\end{eqnarray}
The resulting form of the mass matrix depends on the correlation among 
$\eta_i$. The mass matrices used in the main part are corresponding to the 
following three classes of uniformity:

[Uniform: $\eta_1=\eta_2=\eta_3$]
\begin{eqnarray}
 {\cal M}^{\nu}
 &=&
 \left(\begin{array}{ccc}
 2B & -B & -B \\
 -B & A+B & -A \\
 -B & -A & A+B
 \end{array}\right)\nonumber \\
 &=&\frac{B}{2}
 \left(\begin{array}{ccc}
 4 & -2 & -2 \\
 -2 & 1 & 1 \\
 -2 & 1 & 1
 \end{array}\right)
 +\left(A+\frac{B}{2}\right)
 \left(\begin{array}{ccc}
 0 & 0 & 0 \\
 0 & 1 & -1 \\
 0 & -1 & 1
 \end{array}\right)\ ,
\end{eqnarray}

[Semi-uniform: $\eta_2 = \eta_3$, $\sum\eta_i=0$]
\begin{eqnarray}
 {\cal M}^{\nu}
 &=&
 \left(\begin{array}{ccc}
 B/2 & B/2 & B/2 \\
 B/2 & A+B & -A \\
 B/2 & -A & A+B
 \end{array}\right)\nonumber \\
 &=&\frac{B}{2}
 \left(\begin{array}{ccc}
 1 & 1 & 1 \\
 1 & 1 & 1 \\
 1 & 1 & 1
 \end{array}\right)
 +\left(A+\frac{B}{2}\right)
 \left(\begin{array}{ccc}
 0 & 0 & 0 \\
 0 & 1 & -1 \\
 0 & -1 & 1
 \end{array}\right)\ ,
\end{eqnarray}

[Shared-unifrom: $3\eta_1=\sum\eta_i$,
$\eta_i\neq\eta_j$ $(i\neq j)]$  
\begin{eqnarray}
 {\cal M}^{\nu}=C
 \left(\begin{array}{ccc}
 4 & -2 & -2 \\
 -2 & 1 & 1 \\
 -2 & 1 & 1
 \end{array}\right)\ .
\end{eqnarray}

\subsection{Dirac neutrinos}
Next, let us consider the Dirac neutrino case
\begin{eqnarray}
 -{\cal L}_D=\bar{\nu}_{Li} M^D_{ij} \nu_{Rj} + h.c.\ .
\end{eqnarray}
For Dirac neutrinos, in general, the twisted FL symmetry can be 
imposed on the left- and right-handed neutrinos separately as
\begin{eqnarray}
 &&\nu_{Li} \rightarrow \nu_{Li}^{'}=S_{ij}^L\nu_{Lj}+\eta_{Lj}\xi\ ,\\
 &&\nu_{Ri} \rightarrow \nu_{Ri}^{'}=S_{ij}^R\nu_{Rj}+\eta_{Rj}\xi\ .
\end{eqnarray}
Two independent $\mu - \tau$ permutation symmetries 
make the Dirac mass matrix as
\begin{eqnarray}
 M^D=
 \left(\begin{array}{ccc}
 D & -2C & -2C \\
 -2B & -A & -A \\
 -2B & -A & -A
 \end{array}\right)\ ,
\end{eqnarray}
while the translational symmetries lead to the conditions
\begin{eqnarray}
 &&M^D_{ij}\ \eta_{Rj}=
 \left(\begin{array}{ccc}
 D & -2C & -2C \\
 -2B & -A & -A \\
 -2B & -A & -A
 \end{array}\right)
 \left(\begin{array}{c}
 \eta_{R1} \\ \eta_{R2} \\ \eta_{R3}
 \end{array}\right)=0\ ,
\end{eqnarray}
and
\begin{eqnarray}
 &&\eta_{Li}\ M^D_{ij}=
 \left(\begin{array}{ccc}
 \eta_{L1} & \eta_{L2} & \eta_{L3}
 \end{array}\right)
 \left(\begin{array}{ccc}
 D & -2C & -2C \\
 -2B & -A & -A \\
 -2B & -A & -A
 \end{array}\right)=0\ .
\end{eqnarray}
The resulting form of the mass matrix depends on the correlations among
$\eta_{Li}$ and $\eta_{Ri}$ again.
In this letter, we have assumed the uniform translation,  
that is $\eta_{Li}\propto(1,1,1)$ and $\eta_{Ri}\propto(1,1,1)$.
Then, the mass matrix of the Dirac neutrino takes the form 
\begin{eqnarray}
  M^D=C
 \left(\begin{array}{ccc}
 4 & -2 & -2 \\
 -2 & 1 & 1 \\
 -2 & 1 & 1
 \end{array}\right)\ .
\end{eqnarray}

\end{document}